\documentclass[12pt,epsf,mathfrak]{article}
\usepackage{tabularx}
\usepackage{array}
\usepackage{graphics}
\usepackage{epsfig}
\usepackage{amssymb}

\makeatletter
\usepackage{verbatim}

\newcommand{\<}{\langle}
\renewcommand{\>}{\rangle}
\newcommand{\Tr}{\mbox{Tr}\;} 
\newcommand{\nn}{\nonumber}
\newcommand{\msbar}{\overline{\mbox{\scriptsize MS}}}

\newcommand{\ri}{{\rm RI/MOM}}
\newcommand{\bea}{\begin{eqnarray}}
\newcommand{\eea}{\end{eqnarray}}
\newcommand{\beq}{\begin{equation}}
\newcommand{\eeq}{\end{equation}}
\newcommand{\be}{\begin{equation}}
\newcommand{\ee}{\end{equation}}
\newcommand{\bi}{\begin{itemize}}
\newcommand{\ei}{\end{itemize}}

\newcommand{\pdir}{p\kern -5.2pt\raise 0.2ex\hbox {/}}
\newcommand{\vdir}{v\kern -5.75pt\raise 0.15ex\hbox {/}}
\newcommand{\kdir}{k\kern -5.75pt\raise 0.15ex\hbox {/}}
\newcommand{\epsdir}{\epsilon\kern -5.0pt\raise 0.15ex\hbox {/}}
\newcommand{\bvdir}{\bar{v}\kern -5.75pt\raise 0.15ex\hbox {/}}
\newcommand{\Ddir}{D\kern -7.75pt\raise 0.20ex\hbox {/}}
\newcommand{\ldir}{l\kern -5.0pt\raise 0.2ex\hbox{/}}
\newcommand{\varepsdir}{\varepsilon\kern -5.5pt\raise 0.15ex\hbox{/}}

\newcommand{\kkbar}{K^0-{\overline{K^0}}}

\newcommand{\vettx}{{\bf x}}

\def\ct#1{{\cal #1}}
\makeatother

\begin{document}
\thispagestyle{empty}
\begin{flushright}
{\small 
\begin{tabular}{l}
{\tt DESY, 04-103}\\
{\tt FTUV-04-0629}\\
{\tt IFIC/04-38}\\
{\tt Orsay, 04-45}\\
{\tt RM3-TH/04-14}\\
\end{tabular} 
}
\end{flushright}
\vskip 2.2cm\par

\begin{center}
{\par\centering \textbf{\Large $B_K$ from the lattice with Wilson quarks}}\\
\vskip 0.75cm\par
{\par\centering \sc D.~Be\'cirevi\'c$^a$, Ph.~Boucaud$^a$, V.~Gim\'enez$^b$,\\ 
V.~Lubicz$^c$ and M.~Papinutto$^d$}
{\vskip 0.25cm\par}
\end{center}
{\par\centering \textsl{$^a$Laboratoire de Physique Th\'eorique (B\^at.210),
Universit\'e Paris Sud,}\\
\textsl{Centre d'Orsay, F-91405 Orsay-cedex, France.}\\
\vskip 0.3cm\par}
{\par\centering \textsl{$^b$Dep.de F\' \i sica Te\`orica and IFIC, Univ.~de 
Val\`encia,}\\
\textsl{Dr.~Moliner 50, E-46100 Burjassot,
Val\`encia, Spain.}\\
\vskip 0.3cm\par}
{\par\centering \textsl{$^c$Dip.~di fisica, 
Univ.~di Roma Tre and INFN, Sezione di Roma III, }\\
\textsl{
Via della Vasca Navale 84, I-00146 Rome,
Italy.}\\
\vskip 0.3cm\par}
{\par\centering \textsl{$^d$NIC/DESY Zeuthen, Platanenallee 6, D-15738 Zeuthen, 
Germany}\textsf{\textit{.}}\\
\vskip 1.1cm\par }
\begin{abstract}
We report our results for the bag-parameter $B_K$ obtained from the 
quenched simulations on the lattice with Wilson fermions at three 
values of the lattice spacing. 
We implemented the method by which no subtraction of the mixing 
with other four-fermion $\Delta S=2$ operators is needed. 
Our final result, in terms of the renormalisation group 
invariant bag-parameter, is  $\hat B_K=0.96\pm 0.10$. 
\end{abstract}
\vskip 1.2cm
{\small PACS: 11.15.Ha,\ 12.38.Gc,\ 13.25.Hw,\ 13.25.Jx,\ 13.75Lb,\ 14.40.-n.}
\setcounter{page}{1}                                                            
\noindent
 
\renewcommand{\thefootnote}{\arabic{footnote}}
 
\newpage
\setcounter{page}{1}
\setcounter{footnote}{0}
\setcounter{equation}{0}
\section{Introduction}
The parameter characterising the size of the indirect CP violation in the system of neutral kaons,
$\varepsilon_K$, has been accurately measured since long 
ago~\cite{pdg}. The precise theoretical estimate of the
corresponding $\kkbar$ mixing amplitude, however, is still missing due to  
uncertainties in the computation of the matrix element of the operator 
\be\label{eq:q1}  
O^{\Delta S=2} = ({\bar s}^A \gamma_{\mu}(1-\gamma_5)d^A)
({\bar s}^B  \gamma_{\mu}(1-\gamma_5)d^B)= Q_1 + {\cal Q}_1\, , 
\ee 
where $Q_1$ and ${\cal Q}_1$ are respectively the parity conserving and
parity violating part of $O^{\Delta S=2}$. $A$ and $B$ are the color indices. 
The matrix element of the renormalized operator, 
\bea
\label{def1}
\langle \bar K^0 \vert \hat O^{\Delta S=2}(\mu) \vert K^0 \rangle = 
\langle \bar K^0 \vert \hat Q_1(\mu) \vert K^0 \rangle = 
{8\over 3} f_K^2 m_K^2 B_K(\mu)\  ,
\eea
is conveniently parameterised in terms of the bag-parameter 
$B_K$, the measure of the deviation of the matrix element from 
its value obtained in the vacuum saturation approximation (in which $B_K=1$).

Over the past two decades quite an impressive progress in 
computing $B_K$ on the lattice has been made. We now know how to 
renormalize the four-fermion operator $Q_1$ non-perturbatively 
in the RI/MOM~\cite{bibbia} and in the Schr\"odinger functional scheme~\cite{pena}. 
We also know how to relate to $\hat B_K$, the renormalisation scheme invariant 
bag parameter, since the anomalous dimension  is calculated in a number of schemes at NLO in 
continuum perturbation theory~\cite{guido,ciuchini,sint}, the same accuracy 
at which the corresponding Wilson coefficient has been calculated~\cite{nierste}. 
A high statistics computation of $B_K$ with Wilson fermions 
for several values of the lattice spacing $a$ has been performed in ref.~\cite{JLQCD}. 
The preliminary unquenched calculations have been made too~\cite{BMR}. 
However, all the works in which the Wilson quarks were used 
suffer from the potentially large systematic uncertainty that arise
from the large mixing of the operator $Q_1$ with other parity-even 
operators $Q_{2-5}$ which have different naive chiralities.~\footnote{For the explicit 
forms of all the parity even operators, $Q_{1-5}$, see e.g. Section 3.2 of
ref.~\cite{bibbia}.} That feature 
is a consequence of the explicit chiral simmetry breaking induced by the Wilson term in 
the quark action. In other words,
the renormalization pattern of the lattice operator $Q_1$ 
regularized \`a la Wilson is  
\bea\label{eq:q1ren}
\hat Q_1(\mu) = 
Z(a \mu) \biggl[ \ Q_1(a) + 
\sum_{i=2}^5 \Delta_i(a) Q_i(a) \ \biggr] \;,
\eea
where $Z(a \mu)$ is the multiplicative renormalization constant present
also in formulations where chiral simmetry is preserved, while 
$\Delta_{2-5}(a)$ are the mixing coefficients peculiar for the Wilson 
regularization.
The difficulty is not only that one needs to compute the subtraction 
constants $\Delta_{2-5}(a)$ non-perturbatively
but one should also compute them very accurately because  
the lattice regularized bare matrix elements $\langle Q_{2-5}\rangle$ 
are orders of magnitude larger than  $\langle Q_1\rangle$. 
Therefore, even though the subtraction constants are numerically very 
small (see the tables in ref.~\cite{z4}~\footnote{To avoid a notational 
ambiguity we point out that the subtraction constants $\Delta_{12}(a),
\dots,\Delta_{15}(a)$ presented in ref.~\cite{z4} correspond to $\Delta_{2}(a),
\dots,\Delta_{5}(a)$ used in this letter.}), the net effect of the subtractions 
is large. It is clearly desirable to have a method that allows one
to compute the matrix element~(\ref{def1}) without necessity to subtract the 
mixing.

In this letter we use the hadronic Ward identity, proposed in ref.~\cite{bkws}, 
to relate the matrix element of the operator $Q_1$ to the parity violating one, 
${\cal Q}_1$. The latter does not suffer from the spurious mixing and thus 
the problem of mixing with other dimension-six operators is circumvented. 
The price to pay is that one has to compute a four-point correlation 
function where one pion is integrated over all lattice space-time coordinates.
Similar in spirit, but quite different in practice, is the proposal made 
in ref.~\cite{frezzotti} where the chiral rotation has been added to 
the mass term as to kill out the spurious lattice mixings. 
Preliminary study of the $B_K$ parameter by using that method, has been 
presented in ref.~\cite{tassos}.

In sec.~\ref{sec:2} we will briefly recall the basic elements 
of the Ward identity method to compute $B_K$ without
subtractions; in sec.~\ref{sec:3} we present the results for 
the matrix element~(\ref{def1}) for the directly accessible pseudoscalar 
meson masses from which we will extract the $\hat B_K$ parameter; 
in sec.~\ref{sec:4} we briefly conclude.

\section{Strategy\label{sec:2}}

In this section we will briefly recall the main steps involved 
in the extraction of the $B_K$-parameter by using both methods,
with and without subtractions.

\subsection{Standard Procedure: ``with subtractions"}

The standard way to extract the matrix element~(\ref{def1}) proceeds through the 
computation of the correlation functions
\bea\label{eq:corrs}
G_{K_P^0}(t) = \langle K^{0\ \dagger}_P(t) K^0_P(0)\rangle\,,\qquad 
G_{\hat Q_1}(t_x,t_y) = \langle K^{0\ \dagger}_P(t_x) \hat Q_1(0) K^0_P(t_y)\rangle\,,
\eea
with $\hat Q_1$ defined in eq.~(\ref{eq:q1ren}), and $K^0_P(t_x)=\sum_{\vec x}\bar d(x) \gamma_5
s(x)$. Therefore to get $G_{\hat Q_1}(t_x,t_y)$ one  must compute  
the correlators by using the complete set of parity
conserving four fermion operators, $Q_{1-5}$, subtract the spurious mixing, 
and provide the multiplicative renormalisation, as indicated in
eq.~\ref{eq:q1ren}. 
This procedure is particularly delicate because the approximate restoration 
of chiral symmetry (which is exactly recovered only in the continuum limit) 
depends on how well the subtractions are made. The subtraction 
constants $\Delta_{2-5}$ do not depend on the renormalization scheme. 
Their values have been recently estimated non-perturbatively, in the RI/MOM 
scheme~\cite{z4}.

The matrix element~(\ref{def1}) is extracted from the study of the large time 
asymptotic behaviour of the ratio
\bea\label{ratio0}
 R^{\tiny{\mathrm{stand}}}(t_y) & =& 
\frac{G_{\hat Q_1}(t_x,t_y)}{G_{K^0_P}(t_x) 
G_{K^0_P}(t_y)} \quad {}_{\overrightarrow{\tiny{\textrm{$T\gg t_y\gg T/2$}}}} 
           \quad \frac{\langle \bar K^0
             \vert {\hat Q_1} \vert K^0 \rangle}{ \vert \langle 0 \vert
K^0_P \vert K^0 \rangle \vert ^2 }\,,
\eea
where we fix one of the source operators at $t_x$ so that the kaon state 
which is created by the four fermion operator in the origin is already 
asymptotic when annihilated by $K^0_P(t_x)$. The time $t_y$, instead,  is free. On
the plateaus, $T\gg t_y\gg T/2$, where all the operators are far away from one other, 
we read off the desired matrix element divided by the pseudoscalar density squared.

\subsection{Alternative Procedure: ``without subtractions"}

The method proposed in ref.~\cite{bkws} is based on the use of a Ward identity which  
arise from applying the $\tau_{3}$  axial rotation, 
\bea  
\delta u(x) = i \alpha(x) \gamma_{5} u(x) \, , \quad  \delta \bar u(x) =i
\alpha(x) \bar u(x) \gamma_{5} \, , \nonumber \\   
\delta d(x) = - i\alpha(x)\gamma_{5} d(x) \, , \quad \delta  \bar 
d(x) =  - i \alpha(x) \bar d(x) \gamma_{5} \, , \label{eq:rot} 
\eea
onto the matrix element  $\langle \hat K_{P}^{0}(x) 
\hat {\cal Q}_1(0)\hat K_{P}^{0}(y)\rangle$, where $\hat K_{P}^{0}=Z_{P}K_{P}^{0}$. 
To write down the relevant Ward identity, we introduce the 
bilinear operators
\bea 
K_{S}^{0}(t)= \sum_{\vettx} \bar
d(x) s(x)  \, ,  && \Pi^{0}(x) =  \bar d(x) \gamma_{5} 
d(x)-\bar u (x)\gamma_{5} u(x)\, ,\label{eq:sources} \eea
and the corresponding renormalized $\hat K_{S}^{0}(t)=Z_{S} K_{S}^{0}(t)$.
With these definitions in hands the renormalized Ward identity reads
\bea &\,& 2 \langle \hat  K_{P}^{0}(t_x) \hat Q_{1}(0) \hat K_{P}^{0}(t_y) 
\rangle =  2 m \sum_{z} \langle \hat \Pi^{0}(z) \hat  K_{P}^{0}(t_x) \hat 
\ct{Q}_{1}(0) \hat K_{P}^{0}(t_y)  \rangle \nonumber \\
&\,& -   \langle  \hat  K_{S}^{0}(t_x) \hat 
\ct{Q}_{1}(0) \hat K_{P}^{0}(t_y)  \rangle
-   \langle  \hat  K_{P}^{0}(t_x) \hat 
\ct{Q}_{1}(0) \hat K_{S}^{0}(t_y)  \rangle   +{\cal O}(a) \, ,
 \label{eq:wi1} \eea
where, in view of the fact that we work out of the chiral limit, 
we dropped the sum over the
space-time of the term containing the divergence of the axial current,
which appears, together with the first term on the r.h.s. containing
$2m\Pi^0(z)$. The dropped term is zero when there is no  momentum
transfer between the initial and final states, provided that no 
singularities are encountered. Working at non-zero quark mass then 
ensures that infrared divergences are avoided, so that the integration 
over space-time indeed yields a vanishing value. 
In practice, we work in the SU(3) limit, 
i.e., we take all three quarks to be degenerate in mass, $m_u=m_d=m_s\equiv m$. 
The term on the l.h.s. of eq.~(\ref{eq:wi1}), corresponding to the 
rotation of the operator $\ct{Q}_{1}$, is the desired matrix element. 
The last two terms in eq.~(\ref{eq:wi1}) correspond to the rotation of 
the pseudoscalar kaon sources. These terms, although necessary to saturate the Ward 
identity, disappear in the SU(3) limit as shown in appendix. 
Thus, the Ward identity we use in practice reads 
\be 
\label{eq:final} 
\langle K_{P}^{0}(t_x) \hat  
Q_{1}(0)  K_{P}^{0}(t_y) \rangle  =m(a\mu) \ct{Z}(a\mu) \sum_{z} \langle 
\hat \Pi^{0}(z)   K_{P}^{0}(t_x) \ct{Q}_{1}(0)  K_{P}^{0}(t_y)  
\rangle \equiv G_{\ct{Q}_{1}} (t_x, t_y) \,,
\ee
where we stress the presence of ${\cal O}(a)$ artefacts, i.e., the four fermion 
operators are not improved. Owing to CPS symmetry
the parity-odd operator ${\cal Q}_1$ renormalizes multiplicatively only. 
$\ct{Z}(a\mu)$ has been recently computed non-perturbatively in the  RI/MOM
scheme~\cite{z4}. We use the quark mass, $m(a\mu)=\rho Z_A(a)/Z_P(a\mu)$, defined through the 
non-singlet axial Ward identity,
\be
\label{eq:2rho2}
\rho= \frac {\langle \nabla_0 A_0(t)\, K_P^{0\dagger}(0) \rangle}
{2 \langle K^0_P(t)\, K_P^{0\dagger}(0)\rangle}\;,
\ee
where $A_\mu(t)=\sum_{\vettx}\bar s(x) \gamma_\mu\gamma_5
d(x)$, and $Z_A(a)$ is the axial-current renormalization factor~\cite{z4}. Notice also 
that in eq.~(\ref{eq:final}) the renormalisation constant of the pseudoscalar density, $Z_P(a\mu)$, in 
$m(a\mu)$ cancels against the one in $\hat \Pi^{0}(z)$.
In terms of Feynman diagrams, eq.~(\ref{eq:final}) can be written as
\be 2\Bigl[ C8(t_x,t_y) + D8(t_x,t_y) 
\Bigr] = 2Z_A \rho \Bigl[ CE(t_x,t_y)
+CE(t_y,t_x) +DE(t_x,t_y) + 
DE(t_y,t_x) \Bigr] \, ,
\ee 
where $C8(t_x,t_y)$ and $D8(t_x,t_y)$ correspond to the connected and 
disconnected ``eight" diagrams, while $CE(t_y,t_x)$ and $DE(t_y,t_x)$ refer to the 
connected and disconnected ``emission'' diagrams shown in fig.~1 of ref.~\cite{bkws}. 
Proceeding like in the standard method, the matrix element is extracted from 
the study of the ratio,
\bea \label{eq:ratiows}
R^{\mathrm{w/o\ subtr.}}(t_y) = \frac{G_{\ct{Q}_{1}}(t_x, t_y)}{G_{K^0_P}(t_x) 
G_{K^0_P}(t_y)}\;\;{}_{\overrightarrow{\tiny{\textrm{$T\!\!\gg\!\!
t_y\!\!\gg\!\! T/2$}}}}\;\;\frac{\langle 
\bar K^{0}\vert \hat Q_{1} \vert K^{0}\rangle}{
\vert\langle 0 \vert  K_{P}^{0}\vert K^{0}\rangle\vert^{2}}\,. 
\eea

\section{Extraction of $B_K$\label{sec:3}}

In this section we present our main results. We use both procedures, the 
standard and the one without subtractions, which provides us a useful 
cross-check. Of course the two methods suffer from ${\cal O}(a)$ effects 
that are different in size, but should converge to the same value in the 
continuum limit.

\subsection{Lattices and signals for the ratios~(\ref{ratio0}) and~(\ref{eq:ratiows}) }

The details of our lattice setups were presented in our previous 
publications~\cite{z4,previous}. 
We work at three lattice spacings which correspond to $\beta=6.0$, $6.2$, and to $6.4$.
In each simulation we work with four different values of the quark mass, i.e., 
with four values of the parameter $\kappa$, and compute the correlation functions needed 
to form the ratios~(\ref{ratio0}) and~(\ref{eq:ratiows}). 
\begin{figure}[t!]
\begin{center}
\epsfxsize13.3cm\epsffile{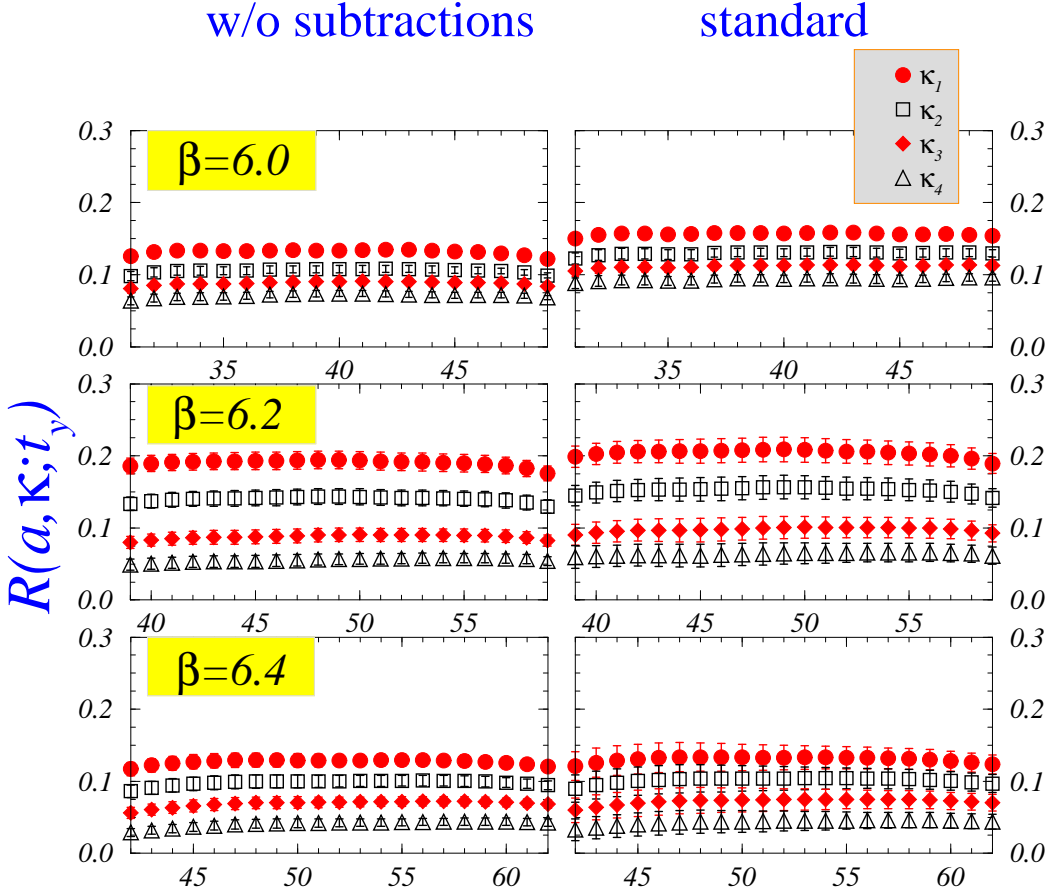} 
\caption{\label{fig1}{ \small \sl Plateaus for the ratios defined in 
eq.~(\ref{ratio0}) and~(\ref{eq:ratiows}) which are referred to as the 
standard procedure (right plots) and the one without subtractions (left plots). 
Plateaus are displayed for all $4$ quark masses and for all three $\beta$'s used in this work.}}
\end{center}
\end{figure}
In fig.~\ref{fig1} we show the signals we obtain by using both methods and for all 
quark masses used in our simulations ($\kappa$'s are ordered as $m_1 > m_2 > m_3>m_4$). 
The plateaus correspond to the signals for the bare operators, i.e. without accounting for 
the overall (scale dependent) renormalisation constants $Z(a\mu)$ and ${\cal Z}(a\mu)$. For the standard method we need to 
specify the subtraction constants $\Delta_{2-5}(a)$. We use the results recently obtained in 
ref.~\cite{z4}. In our calculation one source operator is fixed at  
\bea
t_x =\biggl.\biggl. \biggl.12\biggr|_{\beta=6.0}, 15\biggr|_{\beta=6.2}, 17\biggr|_{\beta=6.4}\,,
\eea
after having checked that the signal does not change for larger $t_x$, except that the  
plateaus become slightly shorter. 
\begin{figure}[h!]
\begin{center}
\epsfxsize8.6cm\epsffile{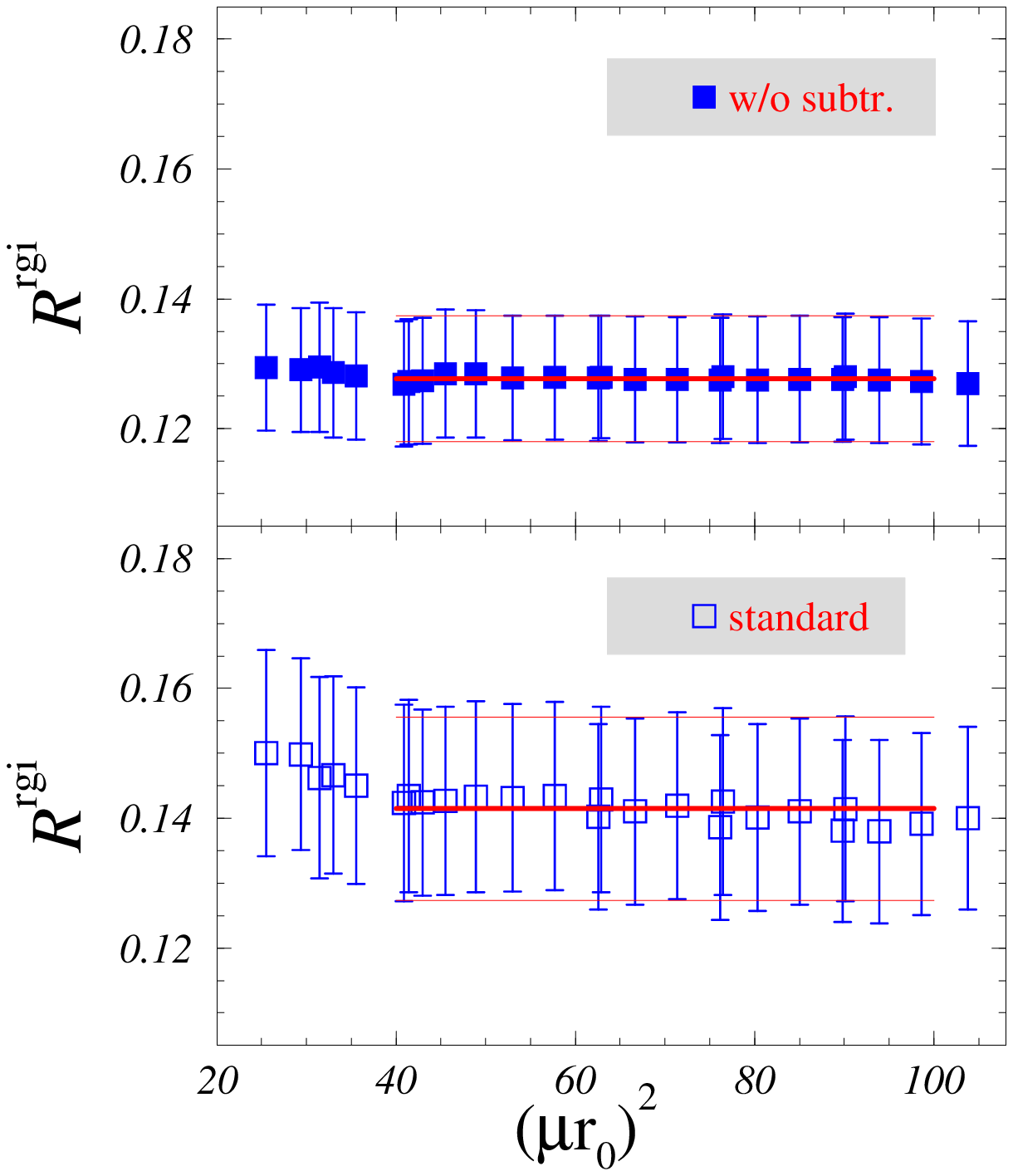} 
\caption{\label{fig2}{\sl \small 
Fit of the ratio $R(t_y)$ [see eqs.~(\ref{ratio0},\ref{eq:ratiows})] to a constant $\hat R\equiv R^{\rm rgi}$ in the interval $40\leq (\mu r_0)^2\leq 100$. 
The lower plot corresponds to the case in which the effect of mixing with other $\Delta S=2$ operators has been 
subtracted. In the upper plot the Ward identity method (without subtractions) has been employed. 
Illustration is provided with the results collected at $\beta=6.2$, and for $\kappa=0.1344$.
}}
\end{center}
\end{figure}
To account for the multiplicative renormalisation we proceed as follows.  
We employ the method described in detail in ref.~\cite{z4}, to compute the renormalisation 
constants $Z(a\mu)$ and ${\cal Z}(a\mu)$ in the RI/MOM scheme at about 20 different 
values of the scale $a\mu$. We then convert such renormalised ratios 
$R^{\rm stand.}(t_y)$ and $R^{\rm w/o\ subtr.}(t_y)$ into their renormalisation invariant forms 
by using the perturbative anomalous dimension known at NLO accuracy~\cite{ciuchini}, namely
\bea
\label{anomalous}
\langle Q_1\rangle^{\rm rgi}    =
\alpha_s(\mu)^{-\gamma_0/2\beta_0}
\left( 1 + {\alpha_s(\mu) \over 4 \pi} J_\ri \right) \ \langle Q_1(\mu)\rangle^\ri\ ,
\eea
where $\gamma_0 = 4$ and $\beta_0 = 11 - 2 n_{\rm f}/3$ are universal and
\bea
J_{\ri} \ = \ 8 \log 2\ -\ {\; 17 397 -2070 n_{\rm f} + 104 n_{\rm f}^2\;
 \over 6 ( 33 - 2 n_{\rm f})^2}\  \,  .
\eea
The plateaus used to fit the ratios $R^{\rm stand.}(t_y)$ and $R^{\rm w/o\ subtr.}(t_y)$ to 
constants $R^{\rm stand.}$ and $R^{\rm w/o\ subtr.}$, respectively, for each value of the 
renormalisation scale $a\mu$, are
\bea\label{plateaux}
t_{\rm plateau}\in \left\{ [36,42]_{\beta=6.0},[43,52]_{\beta=6.2},[49,58]_{\beta=6.4}
\right\}\,.
\eea 
In fig.~\ref{fig2} we illustrate both ratios computed at 24 different values of the renormalisation scale
at $\beta=6.2$, and then converted to the renormalisation group invariant form. After proceeding similarly 
for the other lattice spacings, we find that for $(r_0\mu)^2 \geq 40$, the ratios 
$\widehat R^{\rm stand.}$ and $\widehat R^{\rm w/o\ subtr.}$ nicely follow the perturbatively 
established scale dependence~(\ref{anomalous}). After fitting these results to a constant on the 
interval $40\leq (r_0\mu)^2\leq 100$ we obtain the results listed in table~\ref{tab1}.
To convert from $a\mu$ to $r_0\mu$ we used the accurately estimated $r_0/a$ from ref.~\cite{necco}, while 
in the calculation of the two-loop running coupling, $\alpha_s(r_0\mu)$, we used 
the quenched value $r_0 \Lambda_{\msbar}^{(n_{\rm f}=0)}=0.602(48)$~\cite{capitani}. 
\begin{table}[t!] 
\begin{center} 
\begin{tabular}{|c|c|c|c|c|c|} 
\hline
{\phantom{\Huge{l}}}\raisebox{-.2cm}{\phantom{\Huge{j}}}
{$\beta\quad$}& { $\quad \kappa \quad $ }  & $~ \rho \times 10^{2}~ $  & $\quad X(m_q)\quad $
 & $ \widehat R^{\rm stand.}(a) $ & $\widehat R^{\rm w/o\ subtr.}(a)\quad $  
  \\   \hline \hline 
{\phantom{\Huge{l}}}\raisebox{-.2cm}{\phantom{\Huge{j}}}
$\mathsf{ 6.0}\qquad$   & 0.1335 & 5.997(7)& 0.1256(20) &0.1356(59)& 0.1152(45) \\ 
{\phantom{\Huge{l}}}\raisebox{-.2cm}{\phantom{\Huge{j}}}
                        & 0.1338 & 4.368(8)& 0.1051(19) &0.1124(55)& 0.0925(42) \\ 
{\phantom{\Huge{l}}}\raisebox{-.2cm}{\phantom{\Huge{j}}}
                        & 0.1340 & 3.750(8)& 0.0910(18) &0.0969(52)& 0.0776(39) \\ 
{\phantom{\Huge{l}}}\raisebox{-.2cm}{\phantom{\Huge{j}}}
                        & 0.1342 & 3.129(9)& 0.0765(16) &0.0812(48)& 0.0629(36) \\ \hline
{\phantom{\Huge{l}}}\raisebox{-.2cm}{\phantom{\Huge{j}}}
$\mathsf{ 6.2}\qquad$ & 0.1339 & 5.792(7)  & 0.1766(53)  & 0.188(13)  & 0.1732(97)   \\ 
{\phantom{\Huge{l}}}\raisebox{-.2cm}{\phantom{\Huge{j}}}
                      & 0.1344 & 4.268(7)  & 0.1348(48)  & 0.142(12)  & 0.1277(86)  \\ 
{\phantom{\Huge{l}}}\raisebox{-.2cm}{\phantom{\Huge{j}}}
                      & 0.1349 & 2.748(7)  & 0.0891(40)  & 0.092(12)  & 0.0802(72)  \\ 
{\phantom{\Huge{l}}}\raisebox{-.2cm}{\phantom{\Huge{j}}}
                      & 0.1352 & 1.834(8)  & 0.0589(35)  & 0.060(11)  & 0.0503(57) \\ \hline
{\phantom{\Huge{l}}}\raisebox{-.2cm}{\phantom{\Huge{j}}}
$\mathsf{ 6.4}\qquad$ & 0.1347 & 3.144(2)  & 0.1336(45)  & 0.1363(93)  & 0.1226(79) \\ 
{\phantom{\Huge{l}}}\raisebox{-.2cm}{\phantom{\Huge{j}}}
             	      & 0.1349 & 2.540(3)  & 0.1078(45)  & 0.1087(88)  & 0.0954(74) \\ 
{\phantom{\Huge{l}}}\raisebox{-.2cm}{\phantom{\Huge{j}}}
            	      & 0.1351 & 1.937(3)  & 0.0809(44)  & 0.0804(79)  & 0.0681(65)  \\ 
{\phantom{\Huge{l}}}\raisebox{-.2cm}{\phantom{\Huge{j}}}
            	      & 0.1353 & 1.334(3)  & 0.0536(41)  & 0.0518(65)  & 0.0409(49) \\ \hline
\end{tabular} 
\caption{\label{tab1}
\small{\sl \small The values of the (bare) quark mass, $\rho$, obtained by using the axial Ward identity~(\ref{eq:2rho2}), 
of the quantity $X$, defined in eq.(\ref{XXX}), and of the matrix element $R$ obtained by using both methods. 
$R$ are computed from the fit of the ratios~(\ref{ratio0}) and~(\ref{eq:ratiows})
to a constant on the plateau intervals indicated in eq.~(\ref{plateaux}).}}
\end{center}
\vspace*{-.3cm}
\end{table}
In the same table~\ref{tab1} we also give the values of the bare quark mass $\rho$, computed from the axial 
Ward identity~(\ref{eq:2rho2}), and of the quantity $X(m_q)$ defined as
\be\label{XXX}
\frac{8}{3}\ \frac{Z_A^2 \ \langle  A_0(t)\, A_0^{\dagger}(0) \rangle}{ \langle K^0_P(t)\, K_P^{0\dagger}(0)\rangle}
\quad
{}_{\overrightarrow{\tiny{\textrm{$T\!\gg\!t\!\gg\!0$}}}}
\quad\frac{8}{3} \frac{f^2_P m_P^2}{|\<0|K^0_P|K^0\>|^2}  \equiv X(m_q)\, ,
\ee
where $m_P$ and $f_P$ are the mass and the decay constant of the pseudoscalar meson consisting of 
two degenerate quarks of mass $m_q$.

Before discussing our results for $\hat B_K$, two important remarks are in order. 
In the calculation of $\widehat R^{\rm stand.}$ we used the subtraction constants given in 
ref.~\cite{z4} where beside the statistical we also quoted the systematic uncertainties which 
arise from the spread of values obtained at various values of the momenta flowing through external 
legs of the elementary vertices (i.e., various $a\mu$ in the $\ri$ scheme). In this paper the subtraction and 
renormalisation constants computed in~\cite{z4} are combined with bare matrix elements. The above 
mentioned systematic uncertainties are accounted for by computing the 
renormalised and subtracted matrix element for each value of $a\mu$ separately. Second important remark is 
related to the accuracy of the two methods employed in this paper. From table~\ref{tab1} we see that 
the errors in $\widehat R^{\rm w/o\ subtr.}(a)$ and in $\widehat R^{\rm stand.}(a)$ are comparable.
As mentioned in introduction the computation of the $4$-point correlation function needed for $\widehat R^{\rm w/o\ subtr.}(a)$ 
is more demanding so that --even though one avoids making subtractions-- the statistical errors of the two methods 
are essentially equal. The benefit of the method without subtraction is therefore not in improving  
the statistical quality of the results but rather in preventing the occurence of uncontrollable systematic uncertainties 
that might plague the standard method due to delicate cancellations of subtractions.

\subsection{$\hat B_K$}

With Wilson fermions, ${\cal O}(a)$ lattice artifacts can affect  
the chiral behaviour of the matrix elements relevant to the computation 
of $B_K$. A convenient way for a clean extraction of $B_K$ has been explained in 
ref.~\cite{GAVELA} and consists in studying the dependence of the ratios 
$\widehat R^{\mathrm{stand.}}$ and $\widehat  R^{\mathrm{w/o\ subtr.}}$ as functions of $X$, namely
\bea\label{FITT}
\widehat R \ =\ \alpha\ +\ \beta \ X\,,
\eea
where the fit parameter $\beta$ is identified as $\widehat B_K(a)$, and $\alpha$ is the parameter 
that measures a goodness of the chiral behavior of the ratios $\widehat R$. We find that $\alpha$ 
for all our lattices is consistent with zero. From such fits, at each lattice spacing, we thus 
obtain $\widehat B_K(a)$, all of which are listed in table~\ref{tab2}.
\begin{table}[t!] 
\begin{center} 
\begin{tabular}{|c|c|c|c|} 
\hline
{\phantom{\Huge{l}}}\raisebox{-.2cm}{\phantom{\Huge{j}}}
{$\beta\qquad$}& { $\qquad a/r_0 \qquad $ }  & $\qquad \hat B_K^{\rm stand.}\qquad $  
 & $\qquad \hat B_K^{\rm w/o\ subtr.}\qquad$  
  \\   \hline \hline 
{\phantom{\Huge{l}}}\raisebox{-.2cm}{\phantom{\Huge{j}}}
$\mathsf{ 6.0}\qquad$   & 0.1863 &   1.119(54) & 1.066(39)  \\ 
{\phantom{\Huge{l}}}\raisebox{-.2cm}{\phantom{\Huge{j}}}
$\mathsf{ 6.2}\qquad$   & 0.1354 &   1.074(49) & 1.041(37)  \\ 
{\phantom{\Huge{l}}}\raisebox{-.2cm}{\phantom{\Huge{j}}}
$\mathsf{ 6.4}\qquad$   & 0.1027 &   1.058(44) & 1.017(46)  \\  \hline
{\phantom{\Huge{l}}}\raisebox{-.2cm}{\phantom{\Huge{j}}}
$\mathsf{ \infty }\qquad$ & 0 & 0.980(114) & 0.961(103)  \\ \hline
\end{tabular} 
\caption{\label{tab2}
\small{\small  \sl Results for the $\hat B_K$ parameter as obtained through the fit~(\ref{FITT}) 
for all three values of the lattice spacing and by using both strategies (standard and the 
one without subtractions). The values of $a/r_0$ are taken from ref.~\cite{necco}.
We also show the results of the linear extrapolation in lattice spacing to the continuum limit.
}} 
\end{center}
\vspace*{-.3cm}
\end{table}~\begin{figure}[h!]
\begin{center}
\epsfxsize10.5cm\epsffile{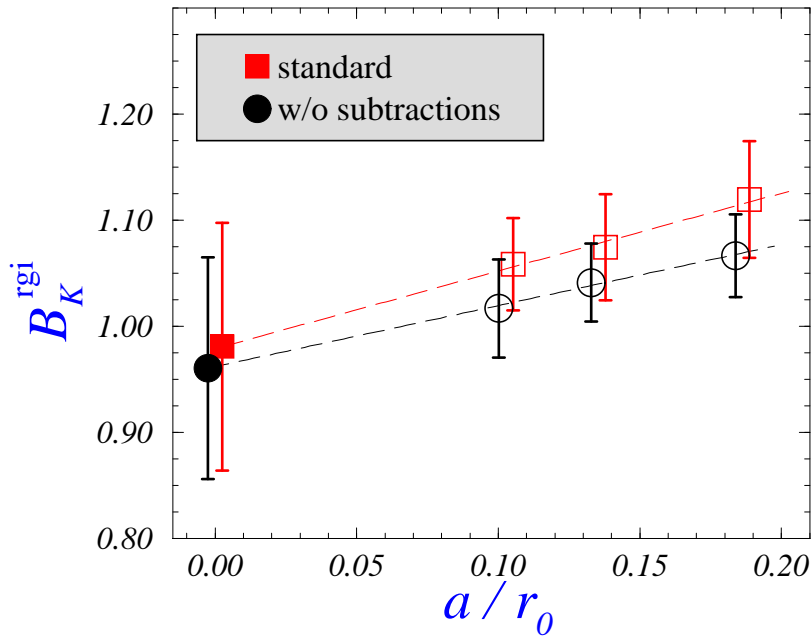} 
\caption{\label{fig3}{\sl \small Extrapolation to the continuum limit. Empty symbols correspond to the 
results obtained at fixed lattice spacing, whereas the filled ones are the results of the linear 
extrapolations. The shapes of the symbols correspond to two different strategies to compute $\hat B_K$, 
as indicated in the legend.}}
\end{center}
\end{figure}
In the same table we also present the results of the extrapolation 
to the continuum limit. That extrapolation has been made linearly since none of the operators used in 
eqs.~(\ref{ratio0},\ref{eq:2rho2},\ref{eq:ratiows}) has been improved. 
We see that the results of the two procedures lead to a consistent value in the continuum limit. 
That may also be viewed as an evidence that no uncontrollable systematic uncertainty used in the standard method 
occured while performing the delicate subtraction procedure.
If we imposed the two methods to produce exactly the same result in the continuum limit 
(similar to what has been done in ref.~\cite{JLQCD-stagger}), we would have obtained
\bea\label{combined}
\hat B_K = 0.969(67)\,.
\eea 
Our errors, after extrapolating to the continuum limit are quite large anyway and we do 
not attempt to include the quadratic term in the continuum extrapolation.
The physical volume of all our lattices is about $(1.7~{\rm fm})^3$. By using the formulae of 
ref.~\cite{fvol} it turns that for the pseudoscalar mesons consisting of degenerate quarks and 
with mass $m_P\gtrsim 500$~MeV, the finite volume effects are negligibly small. 
In the realistic situation, however, one of the valence quarks 
is the strange one (which we can work with directly on the lattice) and the other is $d$-quark. 
That situation would necessitate the chiral extrapolation, which in the quenched theory 
would suffer from the (divergent) quenched chiral logarithms. 
To assess some uncertainty due to the degeneracy we may take the relative difference between the 
chiral logarithmic part known in the degenerate and non-degenerate case in full ChPT. 
With $\Lambda_\chi=1$~GeV, we obtain that $\hat B_K$ for the kaon with non-degenerate 
quarks would be only $2\%
$ smaller than the one with degenerate quarks.
Finally since 
our calculations are made in the quenched approximation, our result cannot 
make impact on the world average value for $\hat B_K$, which is actually completely 
dominated by the errors due to the use of quenched approximation~\cite{reviews}.~\footnote{A complete
list of results for $B_K$ by using other quark actions with recently updated references 
can be found in ref.~\cite{reviews}.} 
It is worth mentioning that the short distance piece in the unquenched scenario would lead to $\hat B_K$ 
larger by only $1\%
\div 2$\%
 compared to the quenched one. Such an estimate arises after replacing $n_{\rm f}=0$ by 
$n_{\rm f}=4$ in eq.~(\ref{anomalous}) and in $\alpha_s(\mu)$, and by using $\Lambda_{\msbar}^{(n_{\rm
f}=4)}=294^{+42}_{-38}$~MeV~\cite{bethke}.

\section{Conclusion\label{sec:4}}

In this letter we presented the results for the renormalisation group invariant 
bag parameter, $\hat B_K$, computed on the lattice with Wilson quarks. 
Besides the standard  procedure, which requires a delicate subtraction of the 
spurious mixing with other $\Delta S=2$, dimension-six, four-quark operators, we also 
implemented the method based on the use of a Ward identity that allows us to 
avoid the subtraction procedure altogether. 

Our lattice data are produced in the quenched approximation at three values of the 
lattice spacing. At each lattice spacing we use the non-perturbatively 
computed renormalisation and subtraction constants, presented in ref.~\cite{z4}. 
The conversion to the standard renormalisation invariant form is made after checking that 
our data follow the renormalisation scale dependence described by the 
RI/MOM anomalous dimension coefficients known to two-loops in perturbation theory.  
After having extrapolated to the continuum limit we obtain the physically relevant 
results quoted in table~\ref{tab2} and eq.~(\ref{combined}). As our final estimate we chose to quote 
the results obtained using the method ``without subtractions", namely
\bea
\hat B_K = 0.96(10)\,.
\eea

\vspace*{10mm}

\section*{Acknowledgment}
We thank Guido Martinelli for the collaboration in the early stages of this work. 
The work by V.G. has been funded by MCyT, Plan Nacional I+D+I (Spain) under the Grant BFM2002-00568.

\vspace*{17mm}

\section*{Appendix}

In this appendix we show that the last two terms of eq.~(\ref{eq:wi1})
vanish in the $SU(3)$ symmetric limit $m_u=m_d=m_s=m$. 

We will use charge conjugation and $\gamma_5$-hermiticity which, on the quark 
propagators $S_f(x,y;U)$ ($f$ is the flavour and $U$ the background 
gauge configuration), act in the following way:
\bea
\textrm{charge conj. }&{\cal C}&\quad  
S_f(x,y;U) = \gamma_0 \gamma_2 S_f^T(y,x;U^C) \gamma_2 \gamma_0\,,\nn \\
\textrm{hermitian conj. }&{\cal H}&\quad  
S_f(x,y;U) = \gamma_5 S_f^\dag(y,x;U) \gamma_5\,,
\eea
where the superscripts T and $\dag$ indicate respectively the transpose
and the hermitian conjugation on color and dirac indices.  
  
Using these two symmetry properties it is easy to show that the trace of an 
arbitrary number of quark propagators and matrices $\Gamma_i\ 
\in\{\textrm{I},\ \gamma_5,\ \gamma_{\mu},\ \gamma_{\mu}\gamma_5,\
\sigma_{\mu\nu}\equiv\frac{1}{2}[\gamma_{\mu},\gamma_{\nu}],\ \tilde
\sigma_{\mu\nu}\equiv\gamma_5\sigma_{\mu\nu}\}$, computed 
on a gauge configuration $U^C$, is the complex
conjugate of that computed on the gauge configuration $U$, i.e.,
\bea
\Tr[\Gamma_1 S_1(x_1,x_2;U)\Gamma_2 S_2(x_2,x_3;U)\ldots\Gamma_n
S_n(x_n,x_1;U)]\;=\nonumber\\\Tr[\Gamma_1 S_1(x_1,x_2;U^c)\Gamma_2
S_2(x_2,x_3;U^c)\ldots\Gamma_n S_n(x_n,x_1;U^c)]^\ast\,.
\eea
This means that taking the real part of the trace corresponds to the inclusion of
the charge-conjugated configuration $U^c$ in the gauge average. Since the QCD action is 
symmetric under the charge conjugation, the average over $N_{\rm conf.}\to \infty$ will 
contain the average over the configuration $U$ and its charge-conjugated one $U^c$.

Another property needed is easily obtained by using hermitian conjugation
and reads
\bea
\label{eq:chargeconj}
&&\qquad\Tr[\Gamma_1 S_1(x_1,x_2;U)\Gamma_2 S_2(x_2,x_3;U)\ldots\Gamma_n
S_n(x_n,x_1;U)]\ =\nonumber\\
&&\left[\Pi_{i=1}^n {\cal E}(\Gamma_i)\right]\;
\Tr[S_n(x_1,x_n;U^C)\Gamma_n\ldots S_2(x_3,x_2;U^C)
\Gamma_2 S_1(x_2,x_1;U^C)\Gamma_1]
\eea
where ${\cal E}(\Gamma_i)=+1$ for
$\Gamma_i\in\{\textrm{I},\gamma_5, \gamma_\mu\gamma_5\}$ and 
${\cal E}(\Gamma_i)=-1$ for $\Gamma_i\in\{\gamma_\mu,\sigma_{\mu\nu},
\tilde\sigma_{\mu\nu}\}$.

We now  analyze the correlation function of $\ct{Q}_1$ between a scalar and a
pseudoscalar source (since we work in the $SU(3)$ symmetric limit we
will not display the flavour indices):
\bea\label{eq:scalarbkws}
\frac{1}{2}\langle K^0_S(x) \ct{Q}_1(0) K^0_P(y)\rangle
\!\! &\!\!=\!\!&\!\! 
\<\bar d(x) s(x)\ \bar s(0) \gamma_\mu d(0) \bar s(0) 
\gamma_\mu\gamma_5 d(0)\ \bar d(y) \gamma_5 s(y)\rangle =\\
&&\langle \Tr \bigl[ S(x,0;U) \gamma_\mu S(0,x;U) \bigr]
\Tr\bigl[\gamma_\mu\gamma_5 S(0,y;U) \gamma_5 S(y,0;U) \bigr]\nn\\
&&+\,\Tr \bigl[ S(x,0;U)\gamma_\mu\gamma_5 S(0,x;U) \bigr]
\Tr\bigl[ \gamma_\mu S(0,y;U) \gamma_5 S(y,0;U) \bigr]\nn\\
&&-\,\Tr\bigl[ S(x,0;U) \gamma_\mu S(0,y;U) \gamma_5 S(y,0;U)
\gamma_\mu\gamma_5 S(0,x;U) \bigr]\nn\\
&&-\,\Tr\bigl[ S(x,0;U) \gamma_\mu\gamma_5 S(0,y;U) \gamma_5 S(y,0;U)
\gamma_\mu S(0,x;U) \bigr]\rangle_U \;,\nn
\eea
where $\langle\dots\rangle_U$ denotes the average over 
gauge field configurations.

Using eq.~(\ref{eq:chargeconj}), we see immediately that the sum of traces
in eq.~(\ref{eq:scalarbkws}) is equal to the same expression computed 
on $U^C$ times ${\cal E}(\gamma_\mu){\cal E}(\gamma_\mu\gamma_5)=-1$. Thus,
including the charge-conjugated configurations in the gauge average give
identically zero for this correlator. Were we not working 
with degenerate $m_s$ and $m_u=m_d$ masses, these
terms should be exponentially suppressed with respect to the 
kaon contribution in the limit of large time distances, 
because they correspond to the propagation of scalar states.
This point can be explicitly
checked by computing $ \langle  \hat  K_{S}^{0}(t_x) \hat 
\ct{Q}_{1}(0) \hat K_{P}^{0}(t_y)  \rangle$ in the same numerical 
simulation for the other correlation functions appearing in 
eq.~(\ref{eq:wi1}).

\end{document}